\documentstyle{l-aa}

\def\lsim{\lower.5ex\hbox{$\; \buildrel < \over \sim \;$}}
\def\gsim{\lower.5ex\hbox{$\; \buildrel > \over \sim \;$}}

\begin{document}

\thesaurus{13 (13.07.2; 09.03.2)}    

\title{Interstellar Gamma Ray Lines from Low Energy Cosmic Ray Interactions}
\author{Reuven Ramaty
}

\institute{Laboratory for High Energy Astrophysics,
           NASA Goddard Space Flight Center,
           Greenbelt, MD 20771, USA
}
\date{Received ; accepted }

\offprints{R. Ramaty}

\maketitle
\markboth{Ramaty: Interstellar Gamma Ray Lines from Low Energy 
Cosmic Ray Interactions}{}

\begin{abstract}

Evidence for the existence of low energy cosmic rays in the Galaxy
comes from the COMPTEL observations of gamma ray line emission from
Orion, and also from light element abundance data which seem to
suggest a low energy rather than a relativistic Galactic cosmic ray
origin for most of the light elements. The Orion and light isotope
data are more consistent with a composition that is depleted in
protons and $\alpha$ particles than with one which is similar to that
of the Galactic cosmic rays. This low energy cosmic ray phenomenon
appears to be highly localized in space and time in the Galaxy, and
is probably associated with star formation regions similar to Orion. 
                                                                                

\end{abstract}

\section{Introduction}

The interactions of accelerated particles with ambient matter 
produce a variety of gamma ray lines following the deexcitation 
of excited nuclei in both the ambient matter and the 
accelerated particles. Apart from solar flares, nuclear 
deexcitation lines following accelerated particle interactions 
have so far only been observed from the Orion molecular cloud 
complex (Bloemen et al. 1994). The Orion observations, carried 
out with the COMPTEL instrument on the Compton Gamma Ray 
Observatory (CGRO) revealed emission features at 4.44 and 6.13 
MeV, due to deexcitations in $^{12}$C and $^{16}$O. Since there 
are no significant long lived radioactive isotopes of 
nucleosynthetic origin that decay into the excited states of 
these nuclei, the observed lines must be produced 
contemporaneously by large fluxes of accelerated particles 
interacting in Orion. Gamma ray emission in the energy range 
from about 30 MeV to 10 GeV was also observed from Orion with 
the EGRET instrument on CGRO (Digel, Hunter, \& Mukherjee 
1995). However, the fact that these EGRET data do not require 
that the cosmic rays in Orion be enhanced relative to the 
relativistic Galactic cosmic rays observed near Earth, implies 
that the particles which produce the line emission in Orion are 
mostly low energy cosmic rays confined to energies below the 
effective pion production threshold.

The discovery of gamma ray line emission from Orion and the 
implied existence of large fluxes of low energy cosmic rays, 
not only in Orion but probably also elsewhere in the Galaxy, 
has led to renewed discussions on the origin of the light 
elements. It has been known for over two decades that cosmic 
ray spallation is important to the origin of  $^6$Li, $^7$Li, 
$^9$Be, $^{10}$B, and $^{11}$B (Reeves, Fowler, \& Hoyle 1970). 
It was shown that the relativistic Galactic cosmic rays (GCR) 
interacting with interstellar matter prior to the formation of 
the solar system may have produced the observed solar system 
abundances of $^6$Li, $^9$Be and $^{10}$B (Meneguzzi, Audouze 
\& Reeves 1971; Mitler 1972). These cosmic rays, however, 
cannot account for the abundances of $^7$Li and $^{11}$B. It is 
believed that about 10\% of the $^7$Li is produced in the big 
bang and that most of the remaining production is due to 
nucleosynthesis in stars (e.g. Reeves 1994). Recent 
measurements of the boron isotopic ratio, $^{11}$B/$^{10}$B, in 
meteorites yielded values in the range 3.84 -- 4.25 (Chaussidon 
\& Robert 1995) which exceed the calculated GCR value by  a 
factor of about 1.5. The implications of the Orion gamma ray 
observations on the origin of the light isotopes have been 
considered by Cass\'e, Lehoucq \& Vangioni-Flam (1995) and 
Ramaty, Kozlovsky \& Lingenfelter (1996). Unlike the GCR, low 
energy cosmic rays, similar to those which produce gamma ray 
line emission in Orion, could account for the meteoritic B 
isotopic ratio. It is in fact possible that the bulk of the 
light isotopes, except $^7$Li, is produced by low energy cosmic 
rays in the Galaxy. Although $^{11}$B could have been produced 
by neutrino spallation of $^{12}$C in supernovae (Woosley et 
al. 1990), recent B and Be observations in stars of various 
metallicities (Duncan 1995) support a low energy cosmic ray 
origin for $^{11}$B since neutrino spallation is not expected 
to produce much Be. 

The present paper is in large part a review based on a series 
of previous articles dealing with the gamma ray line emission 
from Orion and related subjects (Ramaty 1995; Ramaty, Kozlovsky, 
\& Lingenfelter 1995a,b;1996). After a general discussion, we 
consider the implications of the gamma ray line observations on 
the composition, energy spectrum, energy deposition and energy 
density of the low energy cosmic rays in Orion. We also 
consider the ionization produced by these cosmic rays, and 
briefly review the possible origins of the accelerated 
particles. We then proceed to calculate the expected gamma ray 
line emission produced by low energy cosmic rays in the Galaxy. 
We base this calculation on the close relationship between the 
gamma ray line and light isotope production.

\section{General Considerations}

The gamma ray lines produced by accelerated particle 
interactions can be broad, narrow or very narrow (Ramaty, 
Kozlovsky, \& Lingenfelter 1979). Broad lines are produced by 
accelerated C and heavier nuclei interacting with ambient H and 
He. The broadening of these lines (widths ranging from a few 
hundreds of keV to an MeV) is due to the motion of the 
accelerated heavy particles themselves. Narrow lines are 
produced by accelerated protons and $\alpha$ particles 
interacting with ambient He and heavier nuclei. The broadening 
in this case (widths ranging from a few tens of keV to around 
100 keV) is due to the motion of the heavy targets which recoil 
with velocities much lower than those of the projectiles. Very 
narrow lines result from excited nuclei which have slowed down 
and stopped due to energy losses before emitting gamma rays. 
The broadening of these lines is due only to the bulk motion of 
the ambient medium (widths around a few keV or less for the 
interstellar medium).

There are two distinct processes which can lead to very narrow 
line emission: deexcitation of heavy nuclei embedded in dust 
grains and excited by protons or $\alpha$ particles 
(Lingenfelter and Ramaty 1976), and deexcitation of excited 
nuclei populated by long lived radionuclei. Line emission from 
dust is not discussed in the present article. Dust grains, 
however, may play an important role in the injection and 
acceleration of the low energy cosmic rays which produce the 
gamma ray lines (see below). Long lived radionuclei produced by 
accelerated particle bombardment can stop in ambient gas before 
they decay thereby producing excited nuclei essentially at 
rest. The most important such radionuclei are 
$^{55}$Co($\tau_{1/2}=17.5$h), $^{52}$Mn($\tau_{1/2}=5.7$d), 
$^{7}$Be($\tau_{1/2}=53.3$d), $^{56}$Co($\tau_{1/2}=78.8$d), 
$^{54}$Mn($\tau_{1/2}=312$d), $^{22}$Na($\tau_{1/2}=2.6$y), and 
$^{26}$Al($\tau_{1/2}=0.72$my), all of which can be produced in 
accelerated particle interactions, for example 
$^{56}$Fe(p,n)$^{56}$Co. Unlike the very narrow grain lines 
which are produced almost exclusively by accelerated protons 
and $\alpha$ particles, very narrow lines from long lived 
radioactivity can result from both these interactions and 
interactions due to accelerated heavy nuclei. 

In the following discussion we shall need to make assumptions 
on the composition of both the ambient medium and the 
accelerated particles, on the energy spectrum of the 
accelerated particles, and on the interaction model. As these 
have been described in detail in Ramaty et al. (1996), we only 
give a brief summary here. For the ambient medium we assume a 
solar system composition (Anders \& Grevesse 1989). For the 
accelerated particles we consider six different composition: 
solar system (SS), cosmic ray source (CRS), the ejecta of 
supernovae of 35M$_\odot$ and 60M$_\odot$ progenitors (SN35 and 
SN60), the winds of Wolf-Rayet stars of spectral type WC, and 
pick-up ions resulting from the breakup of interstellar grains 
(GR). The grain case is analogous to the anomalous component of 
the cosmic rays observed in interplanetary space (Fisk, 
Kozlovsky, \& Ramaty 1974; Adams et al. 1991). Interstellar 
neutral atoms that penetrate into the solar cavity are picked 
up by the magnetized solar wind after being ionized by solar UV 
and charge exchange with solar wind protons. Because in the 
frame of the wind the ions acquire considerable energy during 
the pick up process ($m_i V^2/2$ where V is the speed of the 
wind), they form a seed population that is much more easily 
accelerated than the rest of the ambient plasma. For Orion it 
is conceivable that the equivalent incoming matter is 
essentially neutral dust that is broken up by evaporation, 
sputtering or other processes. The assumed GR abundances are SS 
abundances modified by depletion factors (Sofia, Cardelli \& 
Savage 1994). The noble gas (He, Ne, Ar) abundances are set to 
zero and H/O = 2, assuming that the bulk of the H and O are in 
ice. The acceleration of pick up ions in Orion was also 
considered by Ip (1995).

The essential properties of these compositions are the 
following:  SS, CRS and to some extent SN35 have large proton 
and $\alpha$ particle abundances relative to C and heavier 
nuclei. On the other hand, because of prior mass loss for SN60 
and WC, and because H and He are essentially absent in dust, 
the SN60, WC and GR compositions are very poor in protons and 
$\alpha$ particles. There is no Ne in the GR composition. The 
WC composition is dominated by C and O and has some additional 
amount of $^{22}$Ne.

All the calculations are carried out in a thick target model in 
which accelerated particles with given energy spectrum and 
composition are injected into an interaction region where they 
produce nuclear reactions as they slow down due to Coulomb 
interactions to energies below the thresholds of the various 
reactions. Energetically this is the most efficient way of 
producing the nuclear reactions; if the particles are allowed 
to escape at higher energies, then energy that would otherwise 
be available for producing nuclear reactions, is removed from 
the system rendering the process less efficient. Energetic 
efficiency is important for gamma ray line production in Orion 
because even under optimal conditions the deposited energy, and 
the accompanying ionization if the  medium is neutral, are very 
large. A model in which the particles escape from the 
interaction region with negligible energy loss is referred to 
as a thin target model. In a thick target the energy spectrum 
of the interacting particles is flatter (because of the energy 
losses) than their source spectrum; in a thin target the 
interacting particle spectrum is identical to the source 
spectrum. A 'thin target' situation can also arise without 
escape, namely when continuous particle acceleration 
compensates for the energy losses so that the spectrum of the 
interacting particles remains identical to their source 
spectrum.

Because the Coulomb energy losses depend on the charge of the 
particles, these losses reduce the importance for gamma ray 
production of heavy nuclei relative to lighter nuclei. Thus, 
for the compositions poor in protons and $\alpha$ particles 
(i.e. SN60, WC and GR), for which practically all the gamma ray 
production is due to C and heavier nuclei, the ratio of the 
line emission from Ne, Mg, Si and Fe to that from C and O is 
smaller in a thick target than in a thin target. This has an 
observable consequence: because C and O produce lines 
predominantly in the 4--7 MeV region while Ne--Fe produce lines 
at energies below 3 MeV, for identical source spectra and 
compositions, deexcitation line production below 3 MeV relative 
to that above this energy is smaller in a thick target than in a 
thin target. This is relevant for Orion, where, as we shall 
see, emission in the 1--3 MeV was strongly suppressed relative 
to the emission between 4--7 MeV.  

The calculations that we present are carried out with two 
spectral forms for the accelerated particle source function, 
\begin{eqnarray}
{dN_i \over dt}(E) = K_i \Big({E \over E_0} \Big)
   ^{-1.5}e^{-E/E_0}, 
\end{eqnarray}
\noindent and 
\begin{eqnarray}
{dN_i \over dt}(E) = K_i \Big({E \over E_c} \Big)^{-s} ~~{\rm 
for}~~ E>E_c \nonumber\\
= K_i ~~{\rm for}~~ E \le E_c,
\end{eqnarray}
\noindent where the $K_i$'s are proportional to the accelerated 
particle source abundances, $E$ is energy per nucleon, and the 
parameters $E_0$ and $E_c$ are allowed to vary over the broad 
range from 2 to 100 MeV/nucl. The nonrelativistic spectral 
index of 1.5 in Eq.~(1) is the consequence (e.g. Ellison \& 
Ramaty 1985) of shock acceleration with a compression ratio at 
its maximal value of 4; the exponential turnover characterizes 
the effects of a finite acceleration time or a finite shock 
size. There is little theoretical basis for the flat spectrum 
given by Eq.~(2). It is a simple form that has been used in 
previous calculations of gamma ray line and light element 
production (e.g. Ramaty et al. 1995a; Cass\'e et al. 1995; 
Ramaty et al. 1996), and we use it here as well; as in the 
previous studies (Ramaty et al. 1995a; 1996) we take $s$ = 10. 
We refer to the spectrum given by Eq.~(1) as the strong shock 
spectrum and to that given by Eq. (2) as the flat spectrum. 

\begin{figure}
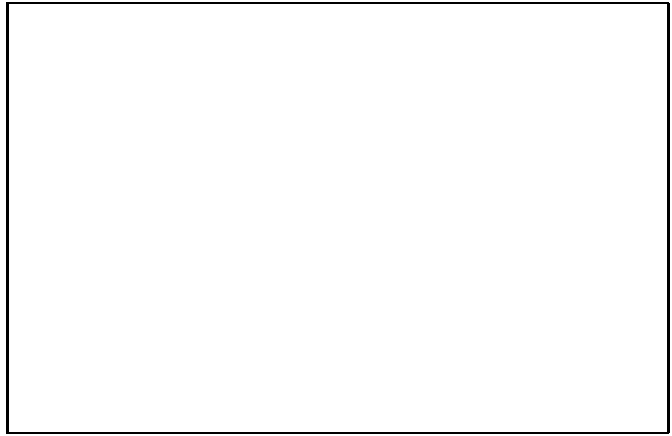

\picplace{5.7 cm}
\caption{Gamma ray observations from Orion; below 30 MeV -- 
COMPTEL data (Bloemen et al. 1994); above 30 MeV -- EGRET data 
(Digel et al. 1995).}
\end{figure}

\section{Gamma Ray Line Emission from Orion}

Gamma ray line emission in the 3 to 7 MeV range was observed 
from the Orion complex with COMPTEL (Bloemen et al. 1994). We 
show the COMPTEL data in Fig.~1 together with the higher energy 
gamma ray emission observed from Orion with EGRET (Digel et al. 
1995). The 3--7 MeV observations show emission peaks near 4.44 
and 6.13 MeV, consistent with the deexcitations in $^{12}$C and 
$^{16}$O following accelerated particle interactions. This 
implies that the ambient matter in Orion is undergoing 
bombardment by an unexpectedly intense, locally accelerated, 
population of low energy cosmic rays. At other photon energies 
the COMPTEL observations reveal only upper limits. As we shall 
see, the 1--3 MeV upper limit places strong constraints on the 
composition of the low energy cosmic rays. Digel et al. (1995) 
have shown that the EGRET data can be understood in terms of a 
relativistic cosmic ray flux similar to that observed near 
Earth producing gamma rays in Orion via pion decay and 
bremsstrahlung. This implies that the low energy cosmic rays 
should be confined to energies below the effective pion 
production threshold (see also Cowsik \& Friedlander 1995). In 
the following we shall be concerned with the properties and 
effects of these low energy cosmic rays.

\begin{figure}
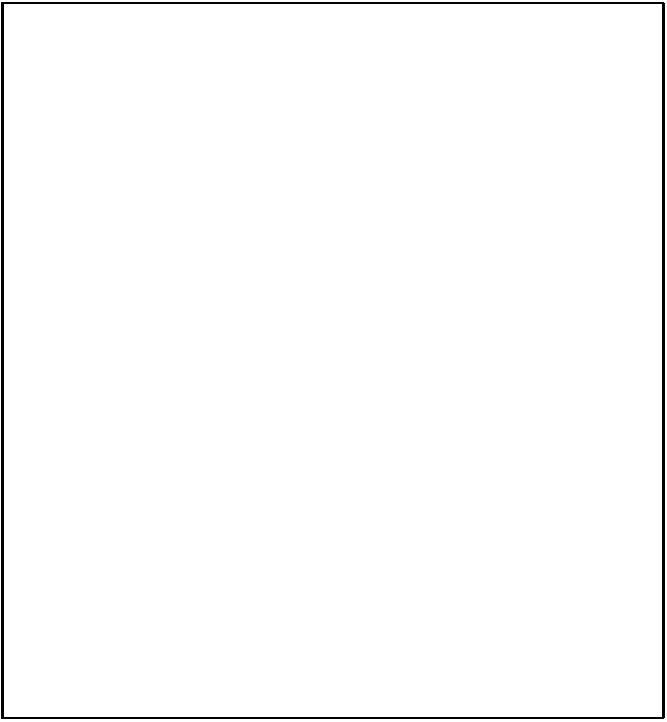

\picplace{9.5 cm}

\caption{Calculated gamma ray spectra for the cosmic ray source 
and the grain compositions, showing a mixed broad-narrow line 
spectrum (upper panel) and a broad line spectrum (lower 
panel); the very narrow lines in the lower panel are from long 
lived radionuclei. The predicted 0.847 MeV line flux in 
the upper panel is 5$\times$10$^{-6}$ ph cm$^{-2}$ s$^{-1}$.}
\end{figure}

\subsection{Composition}

In all the calculations of gamma ray production in Orion the 
ambient gas was assumed to have a solar system composition 
(Anders \& Grevesse 1989). This should not suggest that there 
could be no departures from solar abundances in the ambient 
medium; such effects were simply not yet investigated. The 
results discussed here, therefore, pertain only to the 
accelerated particles.

Information on the proton and $\alpha$ particle abundances 
relative to those of C and heavier nuclei could be obtained 
from observations of the shapes of the gamma ray lines. In 
Fig.~2 we show theoretical gamma ray spectra for the CRS 
(cosmic ray source) and GR (grain) compositions calculated 
using Eq.~2 (flat spectrum) with $E_c$=20 MeV/nucl. (Spectra 
obtained with the strong shock spectrum, Eq.~(1), are not very 
different.) The CRS spectrum (top panel) shows both broad and 
narrow lines, as well as very narrow lines from long lived 
radionuclei. The $^{12}$C complex around 4.44 MeV clearly shows 
the narrow line superimposed on its broad counterpart. On the 
other hand, the narrow lines are absent in the GR spectrum 
(bottom panel) showing the effects of the absence of protons 
and $\alpha$ particles. However, there are still very narrow 
lines from the long lived radionuclei. To allow the shortest 
lived radionucleus $^{55}$Co($\tau_{1/2}=17.5$h) to stop before 
it decays, we assumed that the ambient density exceeds $2 
\times 10^6$ cm$^{-3}$. By convolving spectra similar to those 
in Fig.~2 with Gaussians representing the COMPTEL energy 
resolution, Ramaty et al. (1995a) showed that the current data 
cannot yet rule out a mixed broad-narrow line spectrum (e.g. 
the CRS) even though the COMPTEL energy resolution at 4.44 MeV 
(FWHM$\simeq$300 keV, Sch\"onfelder et al. 1993) would appear 
sufficient. A similar result was also obtained by Cowsik \& 
Friedlander (1995).

\begin{figure}
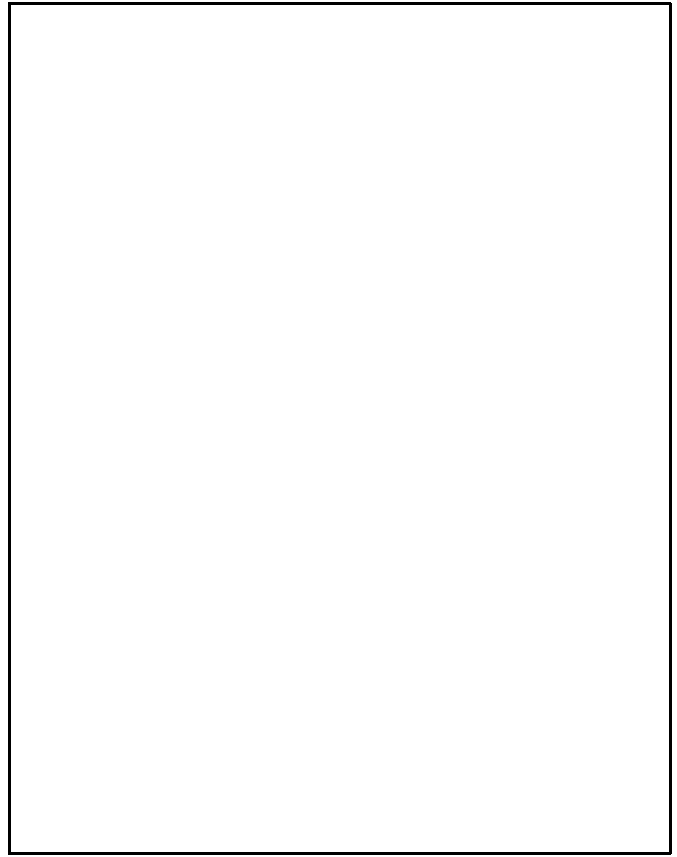

\picplace{11.3 cm}
\caption{Deposited power that accompanies the production of the 
observed gamma ray line emission from Orion for the various 
compositions and energy spectra discussed in the text. The 
ambient medium is neutral and has solar system composition.}
\end{figure}

On the other hand, arguments of energetics tend to support an 
accelerated particle composition poor in protons and $\alpha$ 
particles. To demonstrate this, we calculate the deposited power 
that is associated with the production of the nuclear reactions 
in a thick target. It is given by

\begin{eqnarray}
{dW \over dt} = \sum_{i} 
A_i \int_0^{\infty} E {dN_i \over dt}(E) dE~, 
\end{eqnarray}

\noindent where the $A_i$ is atomic number and ${dN_i \over 
dt}$ is given by Eqs.~1 or 2.  We emphasize that in a thick 
target, the ratio of the nuclear reaction rate to the deposited 
power is essentially independent of details of the interaction 
region, especially the density of the ambient gas. It depends 
mainly on the spectrum and composition of the accelerated 
particles. The composition of the ambient gas could play a 
major role but only if there were a substantial enhancement of 
the C and heavier element abundances relative to H and He. Such 
an enhancement would reduce the deposited power. There is also 
a weak dependence on the state of ionization of the ambient 
gas, as the Coulomb energy losses are higher in a plasma (by 
about a factor of 2) than in a neutral gas. The results are 
shown in Fig.~3 for an ambient neutral medium. The top panel is 
for the strong shock spectrum (Eq.~1) and the bottom panel is 
for the flat spectrum (Eq.~2). The deposited power, normalized 
to the distance of Orion and the observed 3--7 MeV nuclear 
deexcitation flux (Bloemen et al. 1994), is shown as a function 
of $E_0$ or $E_c$ for the various compositions. We see that the 
gamma ray line production is energetically most efficient for 
compositions that are poor in protons and $\alpha$ particles 
(i.e. the SN60, WC and GR compositions), favoring such 
compositions. However, as we shall see below, even in the most 
favorable cases, the rate of ionization of gas in Orion is 
extremely large. 

We mention here that if the light isotopes are indeed produced 
mostly by low energy cosmic rays, then these cosmic rays must 
also be depleted in protons and $\alpha$ particles. The 
$\alpha$ particle depletion is necessary in order not to 
overproduce $^6$Li; the proton depletion ensures a linear 
dependence of the Be and B abundances on the Fe abundance in 
stars of various ages (Duncan 1995). If the low energy cosmic 
rays are poor in protons and $\alpha$ particles they will 
produce Be and B from the breakup of accelerated C and O on 
ambient H and He; in this case both the target and projectile 
abundances could remain constant, leading to a linear growth of 
the Be and B abundances. On the other hand, the GCR would 
produce much of the isotopes from the breakup of C and O in the 
ambient medium whose abundances increase with time, leading to 
a quadratic growth.

Information on the composition of the heavy nuclei can be 
obtained from the observed gamma ray spectrum. In Fig.~4 we 
show calculations of the ratio $R$,
 
\begin{eqnarray}
R = 2 {\int_{1{\rm MeV}}^{3{\rm MeV}} E_{\gamma}^2 
    Q(E_{\gamma}) 
    dE_{\gamma} \over \int_{3{\rm MeV}}^{7{\rm MeV}} E_{\gamma}^2 
    Q(E_{\gamma}) dE_{\gamma}},  
\end{eqnarray}

\noindent for which Bloemen et al. (1994) set a 2$\sigma$ upper 
limit of 0.13 (see also Fig. 1). We see that the CRS and SS 
predicted ratios are in disagreement with the observations by 
more than $3\sigma$, and the SN35 and SN60 predictions are 
inconsistent at greater than $2\sigma$. Modifications in the 
abundances, however, can invalidate these results. For example, 
for the SN60 case, by increasing the C abundance by a factor of 
2, but leaving all the other abundances unchanged, we obtain 
$R\simeq 0.13$ for both $E_0$ or $E_c$ equal 30 MeV/nucl. On 
the other hand, both the GR and WC compositions yield $R$'s 
which are lower than the COMPTEL upper limit. However, while 
the WC composition predicts practically no emission in the 1--3 
MeV region, the GR composition predicts significant broad line 
emission in this region, due to Mg, Si and Fe. The reduction in 
the overall 1--3 MeV emission for the GR case is caused by the 
absence of the 1.634 MeV line due to the lack of Ne in grains. 

\begin{figure}
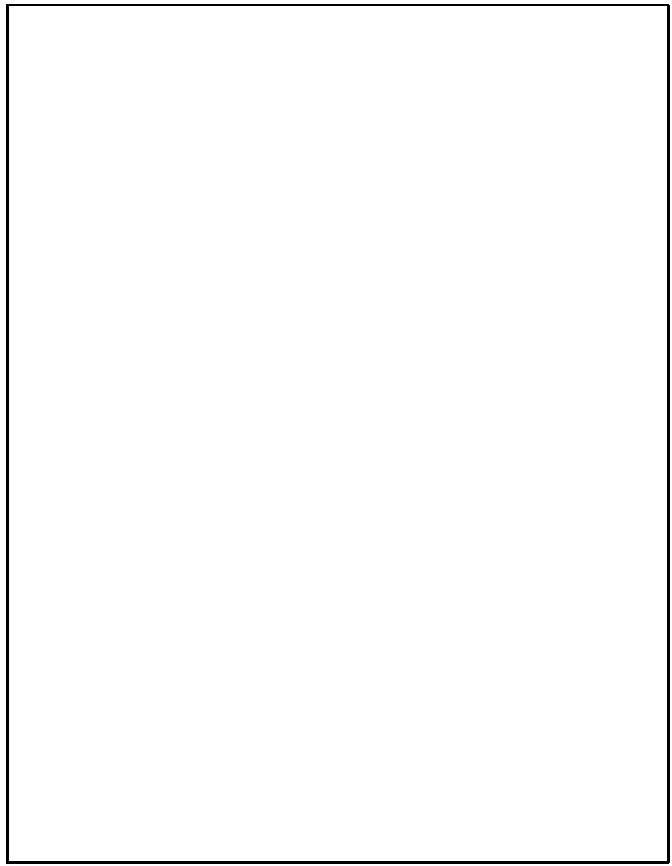

\picplace{11.4 cm} \caption{Calculated values for the ratio $R$ 
defined in Eq.~(4) for various compositions. The upper limit 
established with COMPTEL is plotted at an approximate energy 
which satisfies the constraints imposed by  energetics, line 
shapes and high energy gamma ray observations.}
\end{figure}

\subsection{Energy Spectrum}

We have already pointed out that the fact that the EGRET data 
do not require an enhancement over the locally observed GCR 
implies that the accelerated particles which produce the gamma 
ray line emission in Orion are confined to low energies, below 
the pion production threshold. On the other hand, the 
energetics discussed above require a hard spectrum. This can be 
seen from Fig.~3 which shows that for steep spectra (i.e. small 
$E_0$ or $E_c$) the deposited power becomes very large. Thus 
the combined COMPTEL--EGRET data imply that the low energy 
cosmic rays in Orion should typically have energies around a 
few tens of MeV/nucl. 

The shapes of the observed gamma ray lines could place further 
constraints on the hardness of the accelerated particle 
spectra. For values of $E_0$ or $E_c$ larger than about 50 
MeV/nucleon, for pure broad line spectra, the 4.44 and 6.13 MeV 
lines become very broad (see Ramaty et al. 1996 for details). 
Although the current COMPTEL data cannot yet rule out such 
spectra, future observations could employ line shapes to limit 
the hardness of the low energy cosmic rays.

\begin{figure}
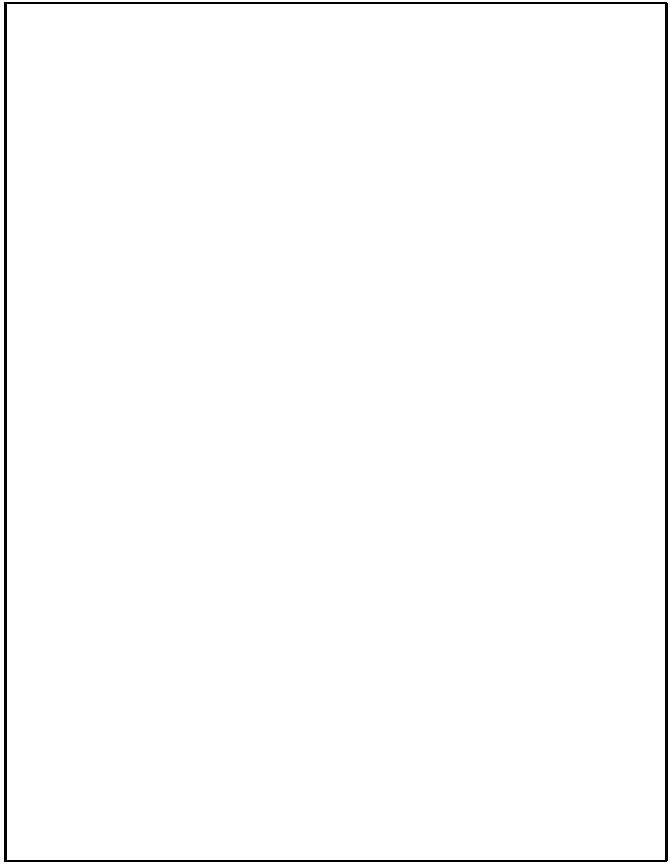

\picplace{11.4 cm} \caption{Energy density in low energy cosmic 
rays in Orion for various values of $E_0$ compared with the 
energy density in the local Galactic cosmic rays.}
\end{figure}

\subsection{Low Energy Cosmic Ray Energy Density}

Whereas the power deposited in Orion by the accelerated 
particles can be calculated independent of the total irradiated 
mass of ambient gas, the low energy cosmic ray energy density  
does depend on this total mass. We have calculated this energy 
density assuming a steady state in which the accelerated 
particles lose energy by Coulomb collisions in a neutral gas. 
As already mentioned, the energy loss rates in an ionized gas 
are higher by only a factor of 2. 

The results are shown in Fig.~5, where $w(E)$ is the 
differential energy density measured in MeV cm$^{-3}$ 
MeV$^{-1}$. The top (bottom) panel is for the CRS (WC) 
composition. Also shown is the energy density in the local 
Galactic cosmic rays (GCR) evaluated from direct cosmic ray 
observations and assuming a power law in momentum source 
spectrum (Skibo 1993). We see that for the CRS composition the 
low energy cosmic rays energy density exceeds the local GCR 
energy density by almost two orders of magnitude; for the WC 
composition the excess is only about a factor of 10.

\subsection{Ionization of the Ambient Gas}

For the SN60, WC and GR compositions and $E_0$ or $E_c$ about 
30 MeV/nucl, the deposited power is (2.5--5)x10$^{38}$ erg 
s$^{-1}$. The total deposited energy depends on the duration of 
the irradiation. For example, if the irradiation lasts 10$^{5}$ 
years, the total energy requirement would be 
(0.8-1.6)x10$^{51}$ ergs, equal to the total kinetic energy 
output of a supernova. Just such a supernova, occurring 
$\sim$80,000 years ago in the OB association at $l$ = 
208$^\circ$ and $b$ = $-$18$^\circ$, the same direction as the 
center of the gamma ray line source, was suggested by Burrows 
et al. (1993) from analy\-ses of the X-ray emission from the 
Orion-Eridanus bubble (see also Ramaty, Kozlovsky \& 
Lingenfelter 1996). 

As it takes 36 eV of cosmic ray energy to produce an ion pair 
in neutral H, the above deposited power corresponds to an 
ionization rate of (4.3--8.7)$\times$10$^{48}$ H atoms s$^{-1}$ 
or (0.17--0.34) M$_\odot$ yr$^{-1}$. Even with no 
recombination, in 10$^5$ years the amount of H that would be 
ionized is only (1.7--3.4)$\times$10$^4$ M$_\odot$, which is 
significantly smaller than the current neutral H mass in Orion. 
Considering the current ionization rate per H atom, we obtain 
that $\zeta$=(0.5--1.0)$\times$10$^{-13}$M$_5^{-1}$ s$^{-1}$, 
where M$_5^{-1}$ is the irradiated neutral H mass in units of 
10$^{5}$ M$_\odot$. The effects of such a very large ionization 
rate have not yet been examined. It is nevertheless possible 
that a large fraction of the power that accompanies the gamma 
ray production is deposited in ionized gas. 

\subsection{Origin of the Low Energy Cosmic Rays in Orion}

In considering the origin of the low energy cosmic rays in 
Orion, we distinguish the problem of the acceleration from that 
of the injection of the particles into the accelerator. The 
proposed acceleration mechanisms are shock acceleration, due to 
shocks associated with the winds of O and B stars (Nath \& 
Biermann 1994) or shocks produced by colliding stellar winds 
and supernova explosions (Bykov and Bloemen 1994), and 
stochastic acceleration, due to gyroresonance with cascading 
Alfv\'en turbulence in the accretion disk of a black hole 
(Miller \& Dermer 1995). The proposed injection sources are the 
winds of Wolf Rayet stars (the WC composition, Ramaty et al. 
1995a), the ejecta of supernovae from massive star progenitors 
(c.f. the SN60 composition, Cass\'e et al. 1995; Ramaty et al. 
1996), and the pick up ions resulting from the breakup of 
interstellar grains (c.f. the GR composition, Ramaty et al. 
1995b;1996). The suppression of accelerated protons and 
$\alpha$ particles relative to C and heavy nuclei (\S 3.1) 
finds an explanation in stochastic acceleration. This 
mechanism, however, does not predict the suppression of Ne and 
heavier nuclei relative to C and O (\S 3.1). The suppression of 
protons and $\alpha$ particles is relatively easily achieved by 
all the proposed injection processes. Concerning the heavier 
nuclei, Ne--Fe are strongly suppressed in WC, but less so in 
the SN60. The suppression of Ne in the GR composition is 
sufficient to account for the 1--3 MeV COMPTEL upper limit.    

The comparison with solar flares is quite instructive (Ramaty 
et al. 1995;1996). The solar flare gamma ray spectra show much 
higher ratios of 1--3 MeV to 3--7 MeV fluxes than does Orion. 
It was shown (Murphy et al. 1991) that this enhanced emission 
below 3 MeV is, in part, due to the enrichment of the flare 
accelerated particle population in heavy nuclei. Such 
enrichments are routinely seen in direct observations of solar 
energetic particles from impulsive flares (e.g. Reames, Meyer 
\& von Rosenvinge 1994). These impulsive flare events are also 
rich in relativistic electrons. On the other hand, in gradual 
events the composition is coronal and the electron-to-proton 
ratio is low. The acceleration in impulsive events is thought 
to be due to gyroresonant interactions with plasma turbulence 
while in gradual events it is the result of shock acceleration. 
The fact that the ratio of bremsstrahlung-to-nuclear line 
emission in Orion is very low lends support to the shock 
acceleration scenario.

\section{Low Energy Cosmic Rays in the Galaxy}

We now provide estimates of the Galactic gamma ray line 
emission expected from low energy cosmic ray interactions. We 
base these estimates on the close relationship between light 
isotope and nuclear deexcitation line production. In Fig.~6 we 
show the B to 3--7 MeV nuclear gamma ray production ratio, 
$Q({\rm B})/Q(3-7)$, as a function of $E_0$ and $E_c$ for the 
strong shock (Eq.~1) and flat (Eq.~2) accelerated particle 
spectra, and for the various compositions. We see that at the 
higher values of $E_0$ and $E_c$ the ratio is not strongly 
dependent on composition. This is because both the B and the 
3--7 MeV deexcitation photons are mostly produced from C and O. 
However, $Q({\rm B})/Q(3-7)$ does depend quite strongly on the 
spectrum of the accelerated particles. For the subsequent 
estimate, we take $Q({\rm B})/Q(3-7)\simeq$0.1, which, for 
$E_0=E_c=20$ MeV/nucl, is a mean for the two assumed spectral 
forms.

\begin{figure}
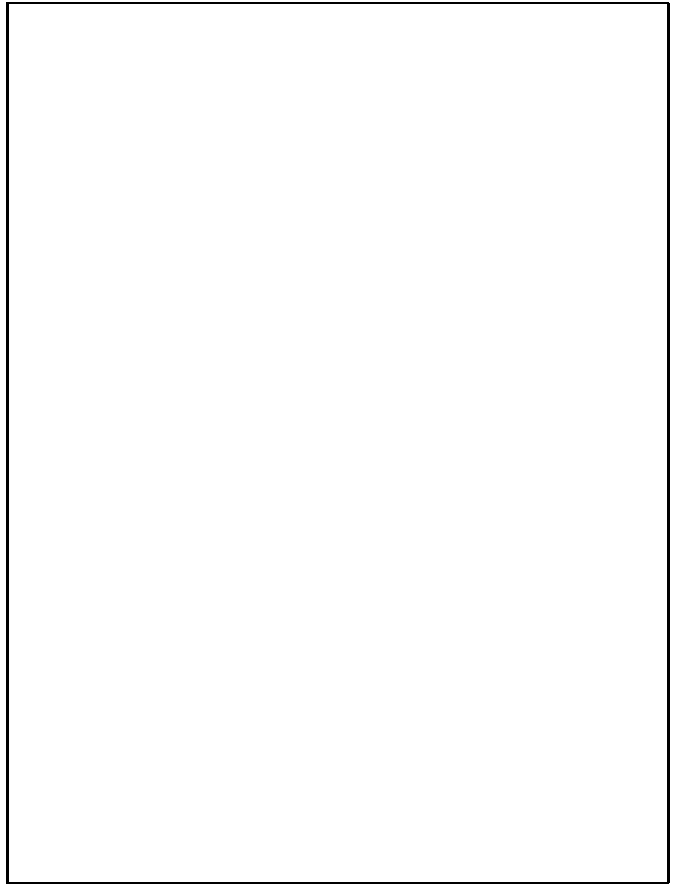

\picplace{11.7 cm} \caption{Production ratios of B to 3--7 MeV 
nuclear deexcitation photons for various abundances and energy 
spectra.}
\end{figure}

To estimate the total B inventory in the Galaxy we first use 
the meteoritic B/H ratio given by Anders \& Grevesse (1989) and 
a total Galactic mass of 5$\times$10$^{10}$ M$_\odot$. This 
yields $N_{\rm B}$$\simeq$3$\times$10$^{58}$ atoms. However, as 
the meteoritic B/H is probably higher than the B/H measured in 
Pop I stars by a factor of 3 to 4 (Reeves 1994), we take the 
Galactic B inventory at the formation of the solar system to be 
approximately 10$^{58}$ atoms, and assume that it was produced 
by low energy cosmic rays in about 5$\times$10$^9$ years. This 
yields an average B production rate prior to the formation of 
the solar system of about 6$\times$10$^{40}$ atoms s$^{-1}$. We 
then further assume that current production rate is equal to 
this average and use $Q({\rm B})/Q(3-7{\rm MeV})\simeq 0.1$ to 
estimate a current Galactic 3--7 MeV photon production rate 
$Q_{\rm G}(3-7{\rm MeV}) \simeq 6\times 10^{41}$ photons 
s$^{-1}$. In terms of this total Galactic production, the flux 
from the central Galactic radian is 

\begin{eqnarray}
\Phi_{3-7}\simeq \xi 10^{-46} Q_{\rm G}(3-7) \simeq 6 
\times 10^{-5}\xi~{\rm ph}~{\rm cm}^{-2}~ {\rm s}^{-1},
\end{eqnarray}

\noindent where $0.5 \lsim \xi \lsim 2$ depending on the 
spatial distribution of the sources (Skibo 1993). Clearly this 
prediction is highly uncertain. The estimate of the current B 
production rate would be larger if there were significant 
destruction of B due to incorporation into stars; in this case 
we would predict a higher central radian flux. On the other 
hand, the current B production rate could be lower than the 
average rate prior to the formation of the solar system, in 
which case our prediction would also be lower. In addition, 
since the B to 3--7 MeV photon conversion depends on the 
spectrum of the accelerated particles, for values of $E_0$ or 
$E_c$ larger than 20 MeV/nucl, the ratio would be smaller than 
the value we used, yielding a lower predicted central radian 
flux. Clearly much could be learned from an actual measurement 
of C and O deexcitation line emission from the direction of the 
Galactic center or elsewhere.

In Fig.~7 we compare our calculations with observations. None 
of these have revealed any line emission. The COMPTEL data 
(Strong et al. 1994) is continuum, most likely a combination of 
bremsstrahlung and inverse Compton radiation produced by 
relativistic electrons. The SMM upper limit refers specifically 
to line emission (Harris, Share, \& Messina 1995). We obtained 
the calculated curves by normalizing the spectra shown in 
Fig.~1 to the 3--7 MeV flux of $6 \times 10^{-5}~{\rm ph}~{\rm 
cm}^{-2}~ {\rm s}^{-1}$ given above. While the calculations 
are not inconsistent with data, it is conceivable that with 
more sensitive instruments or with longer COMPTEL exposures, 
the predicted line emission could be observed.

\begin{figure}
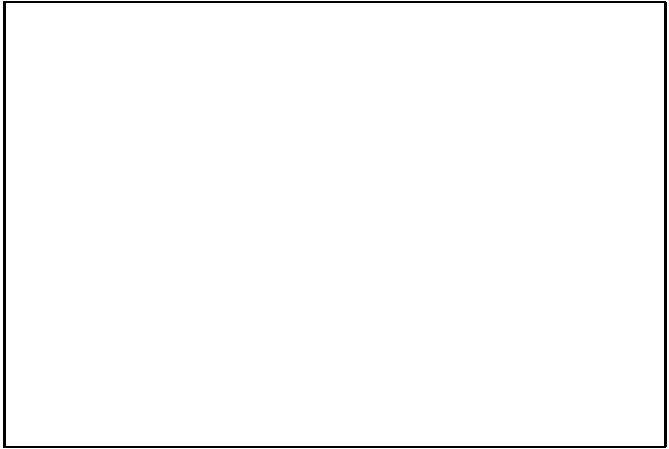

\picplace{5.9 cm} \caption{Predicted nuclear deexcitation line 
emission from the central radian of the Galaxy. The COMPTEL 
continuum observations (Strong et al. 1994) and the SMM upper 
limit on line emission (Harris et al. 1995) are also shown.}
\end{figure}

For the WC, SN60 and GR compositions and $E_0 = E_c = 30$ 
MeV/nucl the energy required to produce 1 B atom is about 1 erg 
(Ramaty et al. 1996). The current rate of B production 
of about 6$\times$10$^{40}$ atoms s$^{-1}$ then implies a 
current total power deposition by low energy cosmic rays in the 
Galaxy of about 6$\times$10$^{40}$ erg s$^{-1}$. When compared 
with the energy deposition rate in Orion (\S  3.4), we find 
that about 200--400 Orion-like regions could be currently 
active in the Galaxy. This small number, and the fact that the 
irradiation time in Orion probably lasted for only 10$^{5}$ 
years, implies that the low energy Galactic cosmic ray 
phenomenon is highly localized in space and time, in contrast 
with the GCR whose spatial distribution is relatively uniform 
and their time dependence, as evidenced by meteoritic studies, 
is relatively constant in time.

\begin{acknowledgements}

I wish to thank S. W. Digel for discussions on the relationship 
between the COMPTEL and EGRET data, and B. Kozlovsky and R. E. 
Lingenfelter with whom much of this research was jointly 
carried out.

\end{acknowledgements}


\begin{thebibliography}{}


\bibitem{} Adams, J. H. et al. 1991, ApJ, 375, L45

\bibitem{} Anders, E., \& Grevesse, N. 1989, Geochim. et Cosmochim. 
Acta, 53, 197.

\bibitem{} Bloemen, H. et al. 1994, A\&A, 281, L5

\bibitem{} Burrows, D. N. et al. 1993, ApJ, 406, 97

\bibitem{} Bykov, A., \& Bloemen, H. 1994, A\&A, 283, L1

\bibitem{} Cass\'e, M., Lehoucq, R., \& Vangioni-Flam, E., 1995, 
Nature, 373, 318 

\bibitem{} Chaussidon, M., \& Robert, F. 1995, Nature, 374, 337

\bibitem{} Cowsik, R. \& Friedlander, M. 1995, ApJ, 444, L29

\bibitem{} Digel, S. W., Hunter, S. D., \& Mukherjee, R. 1995, 
ApJ, 441, 270

\bibitem{} Duncan, D. 1995, paper presented at the Cosmic 
Abundance Conference, College Park, MD, October 1995

\bibitem{} Harris, M. J., Share, G. J., \& Messina, D. C. 1995, 
ApJ, 448, 157

\bibitem{} Ellison, D. C., \& Ramaty, R. 1985. ApJ, 298, 400

\bibitem{} Fisk, L. A., Kozlovsky, B., \& Ramaty, R. 1974, ApJ, 
190, L35

\bibitem{} Ip, W.-H. 1995, A\&A, 300, 283

\bibitem{} Lingenfelter, R. E., \& Ramaty, R. 1976, ApJ, 211, L19

\bibitem{} Meneguzzi, M., Audouze, J., and Reeves, H. 1971, A\&A, 15, 337

\bibitem{} Miller, J. A., \& Dermer, C. D. 1995, A\&A, 298, L13

\bibitem{} Mitler, H. E. 1972, Astrophys. \& Sp. Sci., 17, 186

\bibitem{} Murphy, R. J., Ramaty, R., Kozlovsky, B., \& Reames, D. 
V. 1991, ApJ, 371, 793

\bibitem{} Nath, B. B., and Biermann, P. 1994, MNRAS, 270, L33

\bibitem{} Ramaty, R. 1995, in The Gamma Ray Sky with COMPTON GRO and
SIGMA, eds. M. Signore, P. Salati, and G. Vedrenne, (Dordrecht:
Kluwer), 279

\bibitem{} Ramaty, R., Kozlovsky, B., \& Lingenfelter, R. E. 1979, 
ApJ Suppl., 40, 487

\bibitem{} Ramaty, R., Kozlovsky, B., \& Lingenfelter, R. E. 1995a, 
ApJ, 438, L 21

\bibitem{} Ramaty, R., Kozlovsky, B., \& Lingenfelter, R. E. 
1995b, Ann. New York Academy of Sciences, (17th Texas Symposium 
on Relativistic Astrophysics and Cosmology, eds. H. Bohringer, 
G. E. Morfill and J. Trumper), 759, 392 

\bibitem{} Ramaty, R., Kozlovsky, B., \& Lingenfelter, R. E. 
1996, ApJ, 456, 525

\bibitem{} Reames, D. V., Meyer, J-P., \& von Rosenvinge, T. T. 
1994, ApJ Suppl., 90, 649

\bibitem{} Reeves, H. 1994, Revs. Modern Physics, 66, 193

\bibitem{} Reeves, H., Fowler, W. A., \& Hoyle, F. 1970, Nature, 
Phys. Sci, 226, 727

\bibitem{} Sch\"onfelder, V. et al. 1993, ApJ Suppl., 86, 657

\bibitem{} Skibo, J. G. 1993, Ph.D. Thesis, University of 
Maryland

\bibitem{} Strong, A. W. 1994, A\&A, 292, 82

\bibitem{} Sofia, U. J., Cardelli, J. A., \& Savage. B. D. 1994, ApJ, 
430, 650

\bibitem{} Woosley, S. E., Hartmann, D., Hoffman, R., D., \& Haxton, 
W. C. 1990, ApJ ,356, 272

\end{thebibliography}
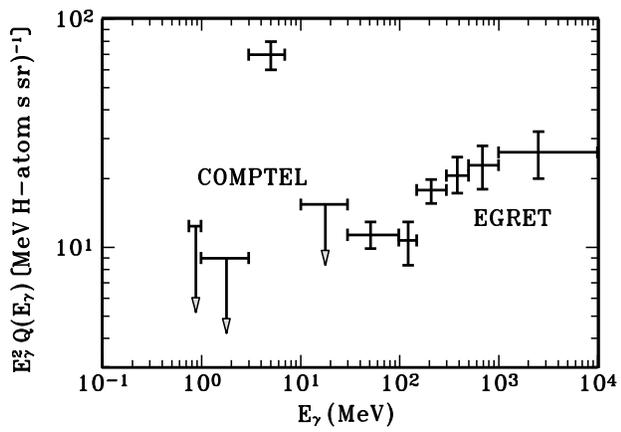
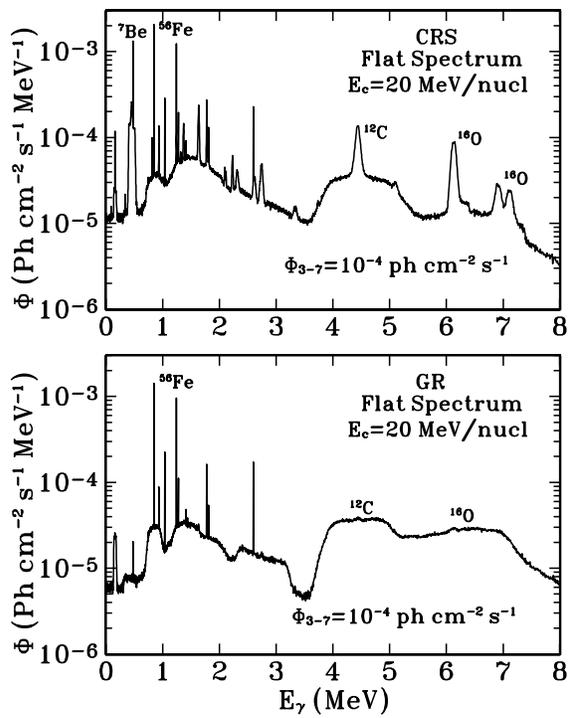
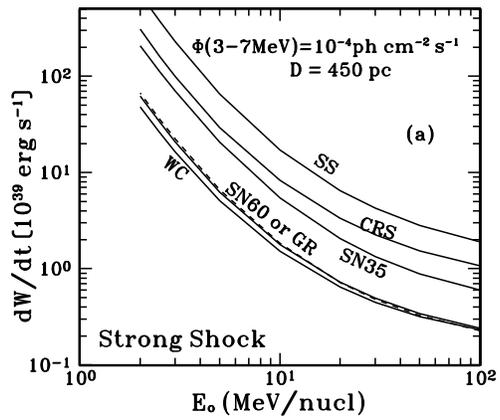
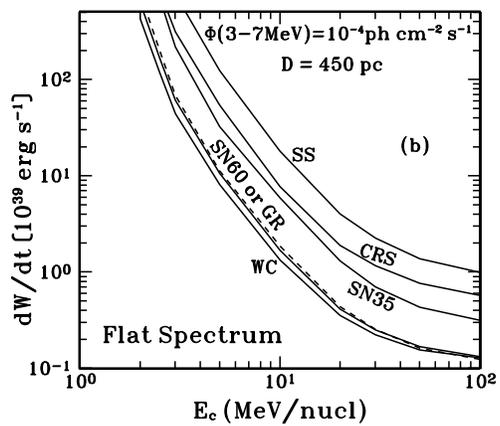
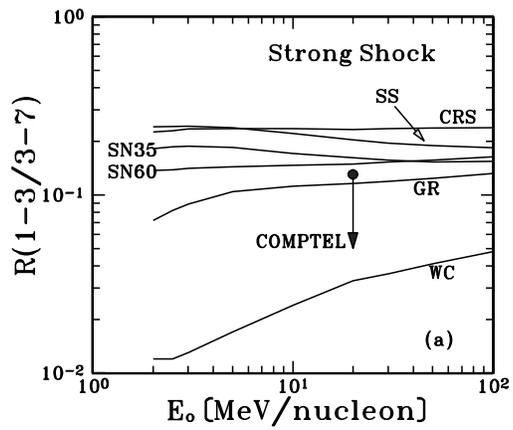
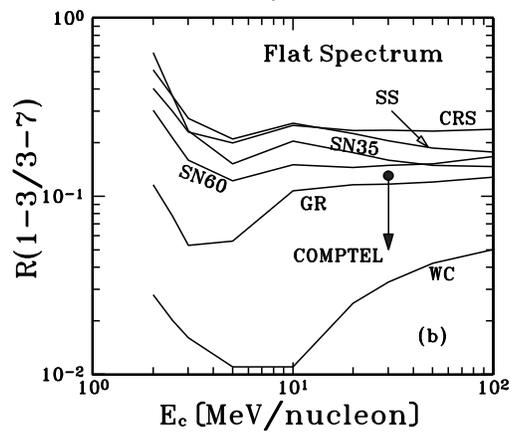
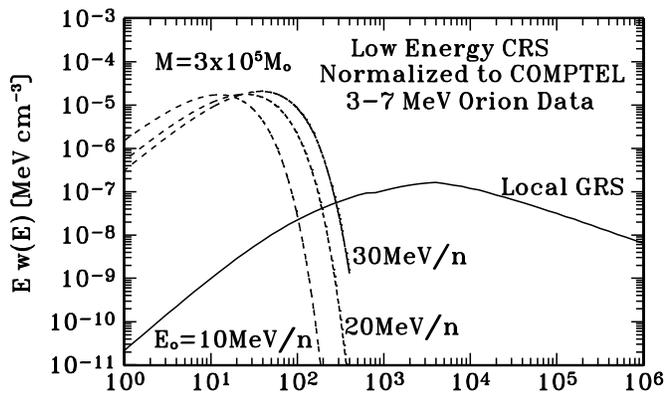
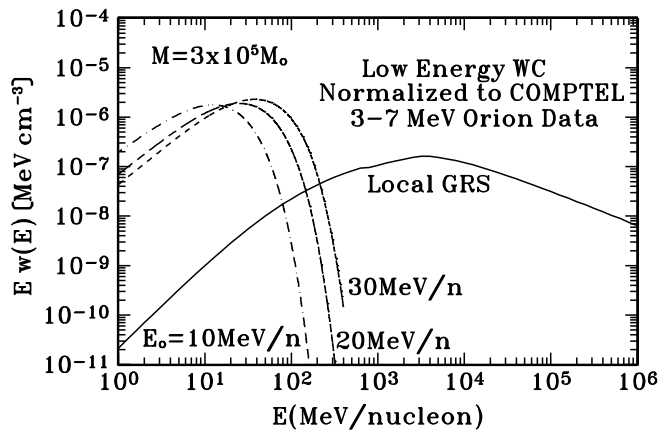
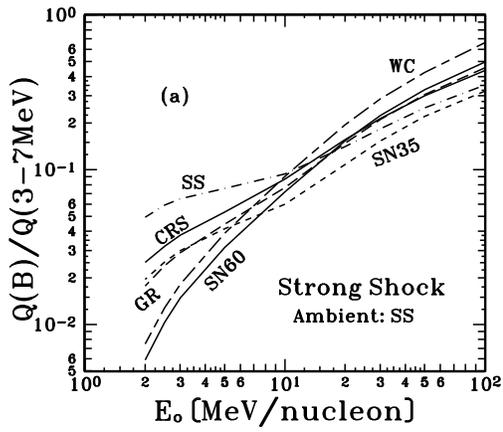
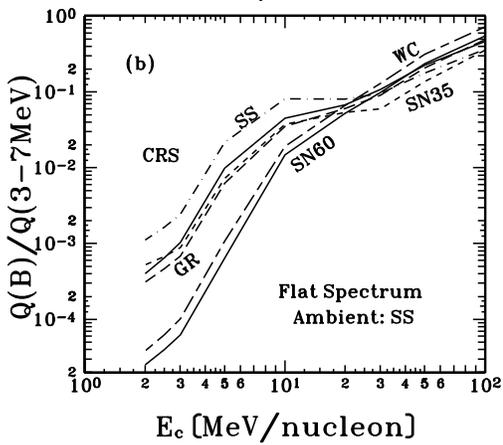
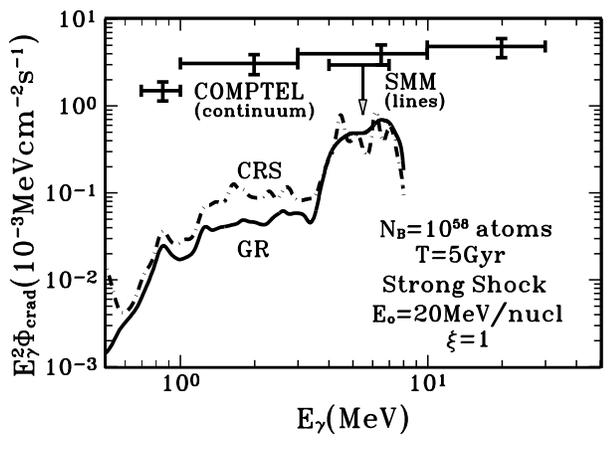
\end{document}